\documentclass[aip,jap,amsmath,amssymb,reprint]{revtex4-1}

\usepackage{graphicx}
\usepackage{dcolumn}
\usepackage{bm}

\usepackage[utf8]{inputenc}
\usepackage[T1]{fontenc}
\usepackage{mathptmx}
\usepackage{etoolbox}
\usepackage{soul}
\usepackage{xcolor}

\makeatother
\begin{document}

\title{Revealing New Leakage Paths in GAAFETs using Predictive First-Principles Simulations}

\author{Denis Mamaluy}
 \email[]{mamaluy@sandia.gov}
\affiliation{Sandia National Laboratories, New Mexico 87123, USA}

\author{Juan P. Mendez}
 \email[]{jpmende@sandia.gov}
 \affiliation{Sandia National Laboratories, New Mexico 87123, USA}

\author{Coleman Cariker}
 \affiliation{Sandia National Laboratories, New Mexico 87123, USA}

 \author{Michael Titze}
 \affiliation{Sandia National Laboratories, New Mexico 87123, USA}
 \affiliation{Luxembourg Institute of Science and Technology (LIST), Belvaux 4422, Luxembourg}

 \author{Alex A. Belianinov}
\affiliation{Sandia National Laboratories, New Mexico 87123, USA}

\date{\today}

\begin{abstract}
Gate-All-Around Field-Effect Transistors (GAAFETs) are emerging as successors to Fin Field-Effect Transistors (FinFETs) and are central to continued scaling at advanced semiconductor nodes. Despite their improved electrostatic control, recent measurements show that the drain current in GAAFETs deviates strongly from thermionic behavior at negative gate voltages, increasing as the gate voltage becomes more negative. This leakage has been attributed to direct gate-to-drain tunneling through the high-$k$ dielectric, enhanced by drain-induced dielectric barrier lowering (DIDBL). Here, we use predictive, charge-self-consistent Non-Equilibrium Green's Function (NEGF) simulations based on the Contact Block Reduction (CBR) method to spatially resolve this leakage pathway in GAAFETs. Unlike approaches based primarily on phenomenological leakage models, our first-principles, multi-terminal quantum-transport framework treats the source, drain, and gate as current-carrying quantum-mechanical contacts and resolves the local current density throughout the device. The spatially resolved current density shows that gate-injected carriers tunnel through the high-$k$ dielectric preferentially near the drain side of the gate, where the dielectric barrier is most strongly lowered by the drain bias, and are collected by the drain. This provides a direct real-space picture of the DIDBL-enhanced gate-drain leakage mechanism.
\end{abstract}

\maketitle

\section{Introduction}

\begin{figure}[t]
    \centering
    \includegraphics[width=1.0\linewidth]{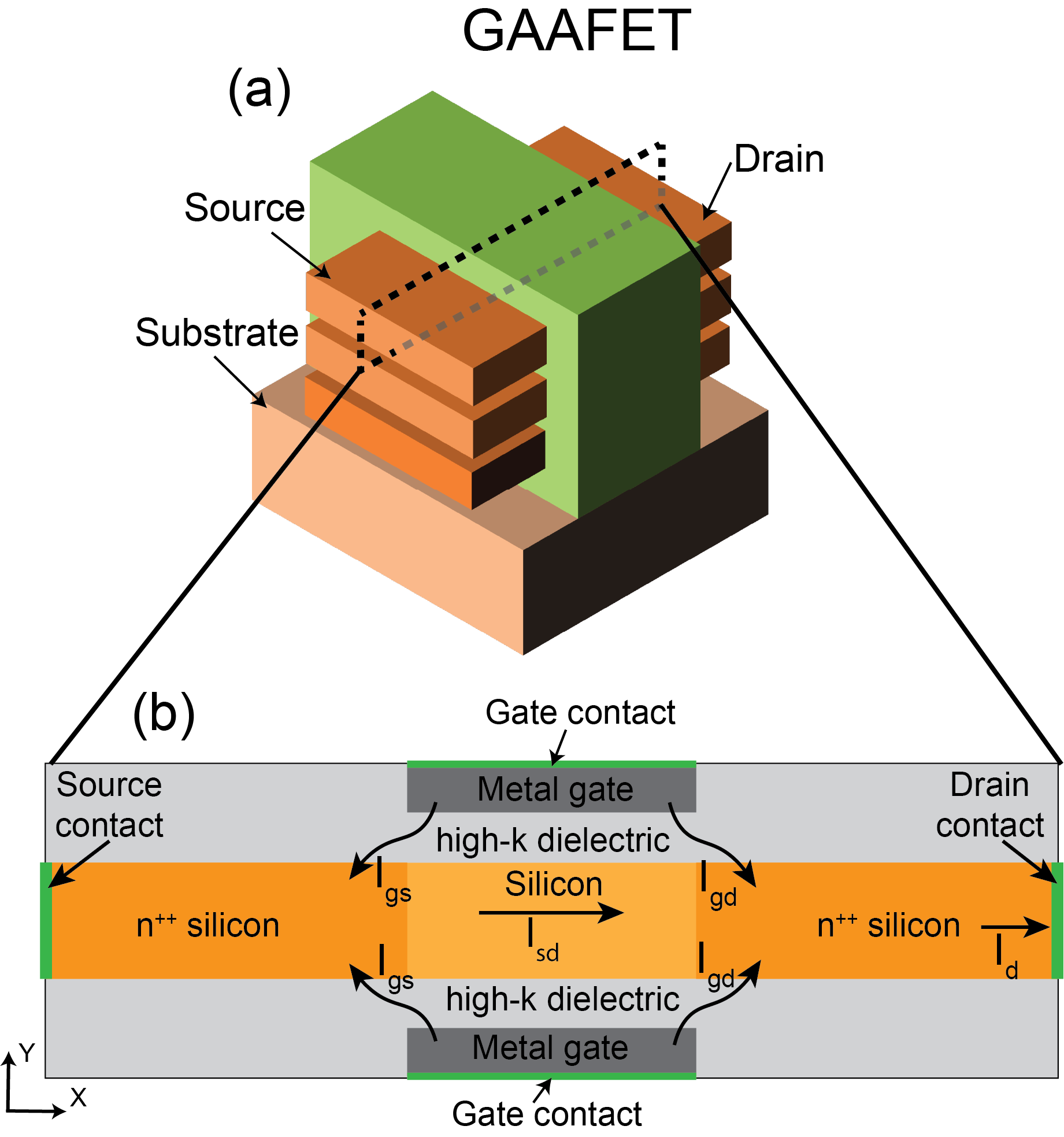}
    \caption{(a) Schematic illustration of the GAAFET architecture. (b) Schematic of the two-dimensional single-sheet GAAFET device model used in this work.}
    \label{fig:device}
\end{figure}

Gate-All-Around Field-Effect Transistors (GAAFETs) are emerging as the leading device architecture for continued CMOS scaling beyond Fin Field-Effect Transistor (FinFET) technology\cite{Loubet:2017,Agrawal:2024,Jaehyun:2023,Eun_Jung:2006,Geoffrey:2024}. By fully surrounding the nanosheet or nanowire channel with the gate (Fig.~\ref{fig:device}), GAAFETs provide improved electrostatic control, reduced short-channel effects, and a pathway toward stacked-channel designs with higher drive current and increased integration density. These advantages make GAAFETs central to advanced semiconductor nodes and future high-performance computing technologies\cite{IEEE_International_Roadmap_for_Devices_and_Systems2023}.

Despite these advantages, recent measurements in Ref.~\onlinecite{Loubet:2017,eipbn2025_4A5} have revealed non-ideal behavior outside the nominal operating regime. In particular, at negative gate voltages, the measured drain current deviates strongly from the expected thermionic trend (see Fig.~\ref{fig:I-V curves}): instead of continuing to decrease exponentially as the channel barrier increases, the current departs from the thermionic trend and rises as the gate voltage becomes more negative. Moreover, other earlier and more recent GAAFET-related studies\cite{Eun_Jung:2006,Jaehyun:2023,Geoffrey:2024,Agrawal:2024} exhibit characteristics that are consistent with this mechanism or, at minimum, do not contradict its presence. Understanding the physical origin of this behavior is important for determining the scaling limits of GAAFETs, reducing off-state leakage, and assessing device reliability in applications where radiation-induced defects predominantly affect the deep-subthreshold regime\cite{Titze:2024}.

In conventional MOSFETs and FinFETs\cite{Chen:1987,Kerber:2013}, drain leakage at negative gate voltage is commonly attributed to gate-induced drain leakage (GIDL), which arises from the strong electric field near the gate-drain overlap region. This strong field induces a significant band bending, enabling band-to-band tunneling and producing an undesired current flow from the grounded substrate to the drain even when the transistor is supposed to be off\cite{Sze:2006,Chen:1987}. However, this mechanism is less applicable to GAAFETs because the active channel is electrically isolated from the substrate by the surrounding high-$k$ dielectric, making a substrate-to-channel-to-drain leakage path unlikely\cite{Mendez:2026}. A second form of GIDL has been reported in ultra-scaled devices, where large negative gate voltages can induce gate-to-channel tunneling followed by band-to-band tunneling \cite{Kerber:2013}. However, the GAAFET design and operating voltage conditions reported in Ref.~\onlinecite{Loubet:2017} are not expected to produce significant band-to-band tunneling. These considerations motivate the investigation of alternative leakage mechanisms using a quantum-transport framework that treats the source, drain, and gate as quantum-mechanical contacts and explicitly accounts for all possible current pathways.

In Ref.~\onlinecite{Mendez:2026}, we identified the physical origin of the negative-gate-voltage leakage in GAAFETs. We showed that the dominant contribution of this leakage current arises from direct gate-to-drain tunneling through the high-$k$ dielectric. This gate-drain leakage observed is further enhanced when the applied drain bias lowers the effective dielectric tunnel barrier between the gate and drain, a mechanism we identified as drain-induced dielectric barrier lowering (DIDBL). In contrast to previous work\cite{Sun:2020,Fan:2015,Sahay:2017,Fan:2013}, which attributed GAAFET leakage to GIDL effect arising from longitudinal band-to-band tunneling occurring between the channel and drain-extension regions and resulting in a source-drain current, our study identified direct gate-to-drain tunneling as the dominant leakage mechanism. This gate-drain leakage mechanism has since been corroborated by experiment\cite{eipbn2025_4A5}. However, while the terminal-current decomposition revealed the dominant leakage component, the spatial pathway of this leakage across the device remained unresolved. In this work, we extend our previous analysis by focusing on the gate-drain leakage current and employing a charge-self-consistent, multi-terminal Non-Equilibrium Green's Function (NEGF) framework implemented using the Contact Block Reduction (CBR) method\cite{Mamaluy:2003,Mamaluy:2005,Gao:2014} to spatially resolve the leakage current and directly visualize how gate-injected carriers tunnel through the dielectric and reach the drain.

\section{Methods}

Simulations use the open-system, charge-self-consistent NEGF formalism\cite{Keldysh:1965,Datta:1997} implemented via the CBR method\cite{Mamaluy:2003,Mamaluy_2004,Sabathil_2004,Mamaluy:2005,Khan:2007,Khan_2008,Gao:2014,Mendez:2021,Mendez:2023}. This framework has previously reproduced experimental results in several nanoscale device systems, supporting its robustness and predictive capability\cite{Commun_Phys:2021,Mendez_CS:2022,Mendez:2023b,Mendez:2026}. The CBR method enables efficient calculation of the density matrix, transmission function, and related quantities for arbitrarily shaped multi-terminal two- and three-dimensional open devices within the NEGF formalism, with computational cost that scales linearly as $O(N)$ with system size ($N$). Unlike conventional two-terminal NEGF implementations\cite{Lake:1997} in which the gate is treated only electrostatically, the multi-terminal NEGF/CBR framework treats the source, drain, and gate as current-carrying quantum-mechanical contacts within the same self-consistent solution. This capability is essential for accurately calculating gate leakage and separating source-drain, gate-source, and gate-drain current components.

Within this framework, the open-system Schr\"{o}dinger equation and the nonlinear Poisson equation are solved self-consistently\cite{Mamaluy:2003,Mamaluy:2005,Gao:2014}. The kinetic energy operator is described using an effective-mass tensor approximation with three non-equivalent, double-degenerate silicon valleys defined by the longitudinal and transverse effective masses.
From the same self-consistent solution, the spatially resolved current density $\mathbf{J}(\mathbf{r})$ is computed using the CBR representation of the nonequilibrium density matrix. The local current between neighboring grid sites is evaluated from the off-diagonal density-matrix elements and the corresponding Hamiltonian couplings, allowing the internal current pathways to be reconstructed and visualized\cite{Mendez:2026b}.

Additionally, in this work, we use Ref.~\onlinecite{Loubet:2017} as the benchmark because it provides one of the most complete available datasets for GAAFETs to date, including device dimensions and measured electrical characteristics. These details are essential for constructing the predictive quantum-transport model and performing a quantitative comparison with experiment.

\section{Results and discussion}

\begin{figure}[t!]
    \centering
    \includegraphics[width=1.0\linewidth]{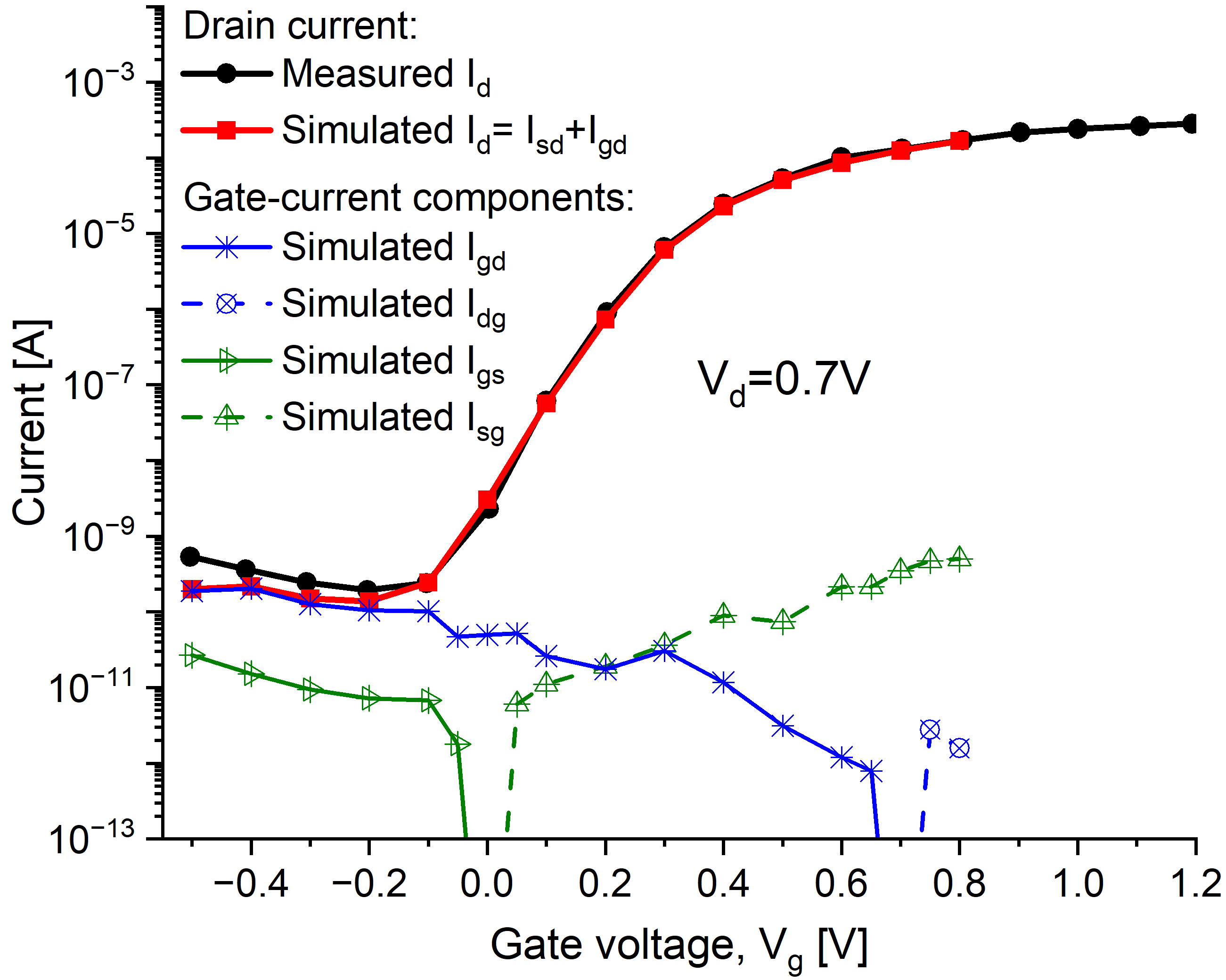}
    \caption{$I-V_{g}$ characteristics and the corresponding gate-leakage current components at $V_d=0.7$~V. All current components are plotted as absolute values, with the current direction indicated by the subscripts. $I_d$, $I_{gd}$, and $I_{gs}$ denote the total drain current, gate-to-drain current, and gate-to-source current, respectively. $I_{dg}$ and $I_{sg}$ denote the drain-to-gate and source-to-gate current components, respectively, and are equal in magnitude but opposite sign to $I_{gd}$ and $I_{gs}$ ($I_{dg}=-I_{gd}$, $I_{sg}=-I_{gs}$). Measured currents from Ref.~\onlinecite{Loubet:2017}. Figure adapted from J.P. Mendez, C. Cariker, M. Titze, A. Belianinov, and D. Mamaluy, J. Appl. Phys. 139, 214501 (2026); licensed under a Creative Commons Attribution (CC BY) license.}
    \label{fig:I-V curves}
\end{figure}

\begin{figure}[t!]
    \centering
    \includegraphics[width=1.0\linewidth]{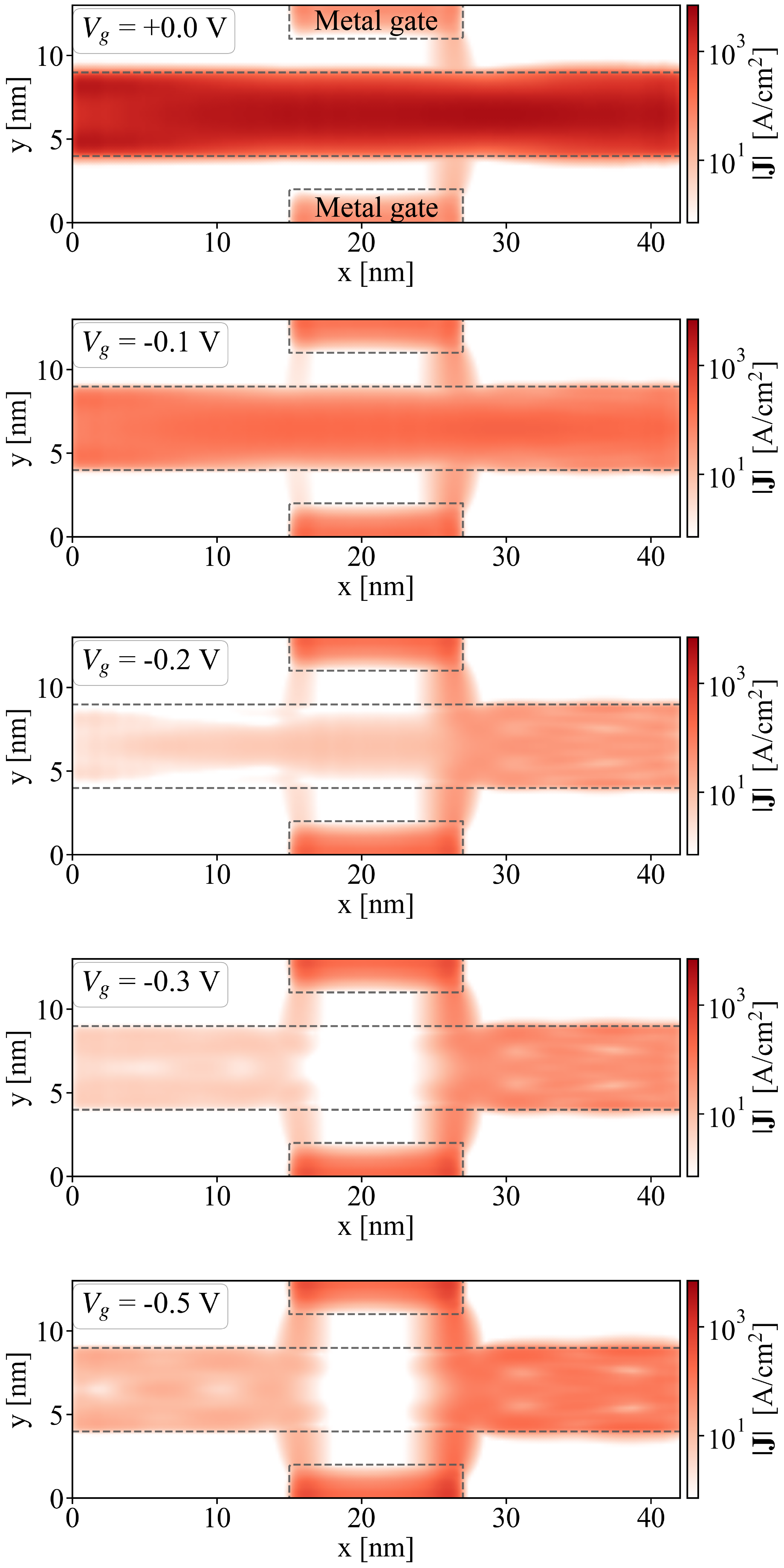}
    \caption{Spatial current density across the device at zero and several negative gate voltages for a drain bias of $V_d=0.7$~V, illustrating the transition from source-drain transport to gate-drain leakage. The magnitude of the current density is computed as $|\bm{J}|=(J_x^2+J_y^2)^{1/2}$. The dash lines indicate the boundaries of the gate, n$^{++}$ silicon source and drain regions, and the channel, as shown in Fig.~\ref{fig:device}}
    \label{fig: Spatial current density}
\end{figure}

\begin{figure}[t!]
    \centering
    \includegraphics[width=1.0\linewidth]{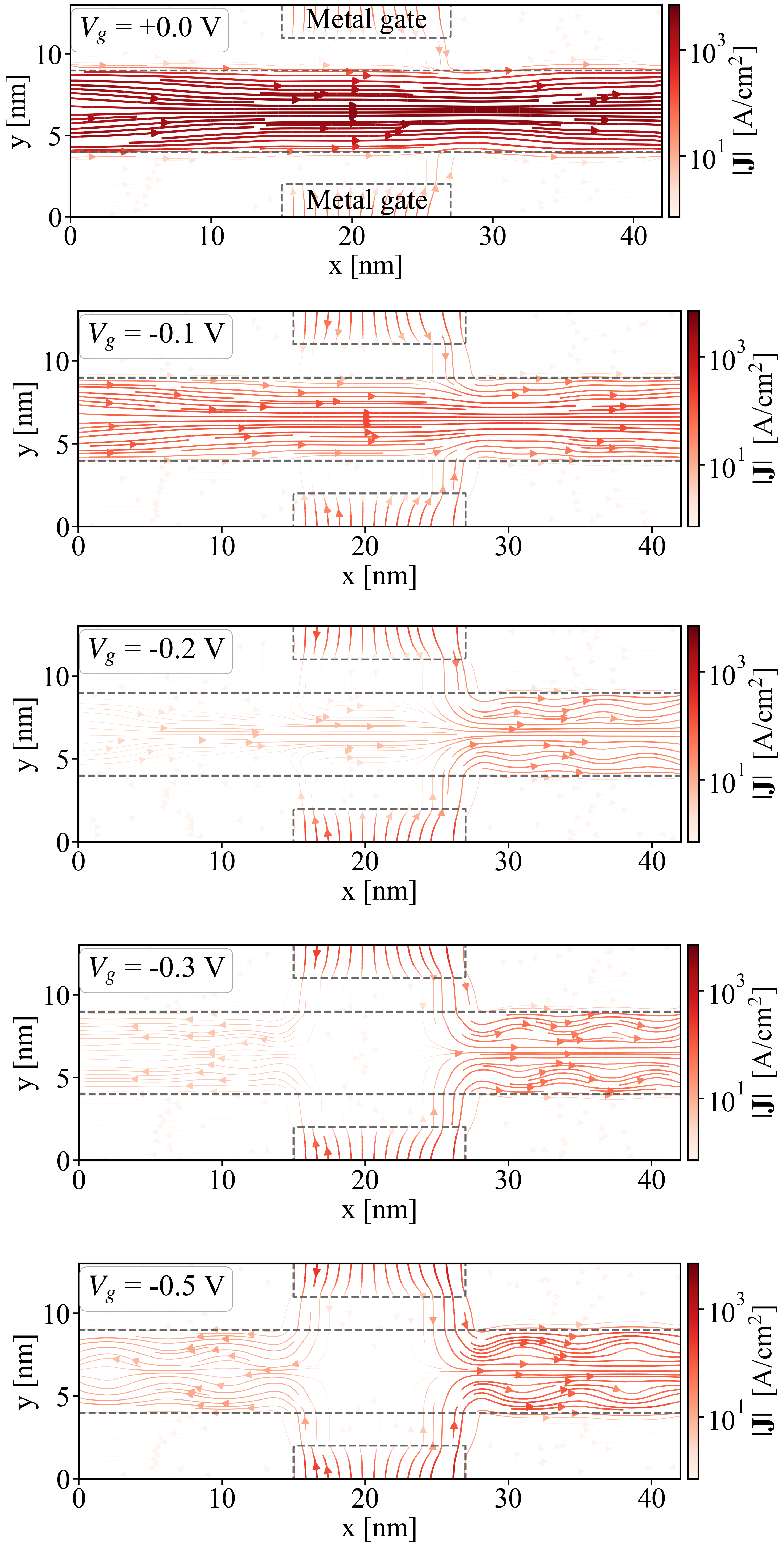}
    \caption{Electron-flow streamlines across the device at zero and several negative gate voltages for a drain bias of $V_d=0.7$~V, illustrating the transition from source-drain transport to gate-drain leakage. The magnitude of the current density is computed as $|\bm{J}|=(J_x^2+J_y^2)^{1/2}$. The current direction is defined by the direction of electron flow. The dash lines indicate the boundaries of the gate, n$^{++}$ silicon source and drain regions, and the channel, as shown in Fig.~\ref{fig:device}}
    \label{fig: Current density streamline}
\end{figure}

For our simulations, we adopt a dual-gate representation shown in Fig.~\ref{fig:device}(b) to investigate the gate leakage in GAAFET structure (Fig.~\ref{fig:device}(a)). The simulated device consists of a two-dimensional (2D) single-sheet GAAFET structure, in which the silicon nanosheet channel is surrounded by a high-$k$ gate dielectric and a metal gate; the third dimension is assumed to be sufficiently large to support a flat potential and plane-wave basis. This simplification is justified by the device geometry, as the nanosheet width is considerably larger than its thickness. Consequently, the electrostatic control and carrier transport are predominantly governed by the top and bottom interfaces, while the contribution from the sidewalls is comparatively small. Therefore, neglecting edge-related effects in our model is expected to introduce only a minor quantitative deviations, without affecting the underlying physical mechanism or the main conclusions of this analysis. Following Ref.~\onlinecite{Loubet:2017}, the physical gate length is L$_g$ = 12~nm, the silicon nanosheet thickness is T$_{Si}$ = 5~nm, and the effective nanosheet width is W$_{Si}$ = 30~nm. The source and drain regions are highly doped with a concentration of $3.0 \times 10^{20}$~cm$^{-3}$, while the channel doping is $5.0 \times 10^{17}$~cm$^{-3}$. The gate stack uses a 2~nm HfO$_2$ high-$k$ dielectric with a dielectric constant of $\epsilon = 20$\cite{Lin:2002} and an electron barrier height of 1.6~eV\cite{HfO2:2000,HfO2}, together with a metal gate work function of 4.5~eV.

Fig.~\ref{fig:I-V curves} shows the measured and simulated current--gate-voltage characteristics at V$_d$ = 0.7~V, together with the decomposed gate-current components, I$_{gd}$ and I$_{gs}$, corresponding to gate-drain and gate-source currents. The simulated total drain current (I$_d$) is in very good agreement with the measurement over the full gate-voltage range. For positive gate voltages, the total drain current is the source-drain channel current, while the gate-current components remain several orders of magnitude smaller. At negative gate voltages, the behavior changes qualitatively: the gate-to-drain current (I$_{gd}$) rises to become the dominant contribution to the total drain current and accounts for its increase as V$_g$ becomes more negative. Another observation from Fig.~\ref{fig:I-V curves} is that the direction of the gate-current components changes with gate voltage. For $V_g>V_d$, electrons flow predominately from the drain to the gate ($I_{dg}$). As the gate voltage is reduced below the drain bias, the direction reverses and electrons flow from the gate to the drain ($I_{gd}$). Similarly, when the gate voltage becomes sufficiently lower than the source bias, electrons flow from the gate to the source ($I_{gs}$) rather than from the source to the gate ($I_{sg}$).

Figs.~\ref{fig: Spatial current density} and~\ref{fig: Current density streamline} present the spatially resolved current density and its flow direction, the central result of this work. Fig.~\ref{fig: Spatial current density} shows the magnitude $|\mathbf{J}(\mathbf{r})|$ and Fig.~\ref{fig: Current density streamline} the corresponding streamlines, oriented along the electron flow, at V$_g$=0~V and several negative gate voltages. At V$_g$=0~V and small negative voltages, the current is confined to the silicon channel and flows directly from source to drain, as expected for normal operation. As the gate voltage becomes more negative, the channel current is progressively suppressed and a qualitatively different pattern develops: carriers are injected from the metal gates, tunnel across the high-$k$ dielectric, and are collected by the drain.

This gate-to-drain leakage path is strongly asymmetric: it forms preferentially at the drain-side edge of the gate, whereas the source-side region of the gate carries comparatively little current. This drain-side localization is the real-space signature of DIDBL, as the drain bias lowers the dielectric barrier most effectively where the gate and the highly doped drain are in close proximity, opening the leakage path precisely there. The streamlines make the mechanism explicit, showing carriers entering from both the top and bottom gates near the drain edge, turning, and converging into the drain. By V$_g$=$-0.5$~V, the source-drain channel path is strongly suppressed and the device current is carried almost entirely by this gate-drain leakage path. These maps thus localize, in real space, the leakage that the terminal-current decomposition could attribute only to the gate-drain terminal pair.

\begin{figure}[t!]
    \centering
    \includegraphics[width=1.0\linewidth]{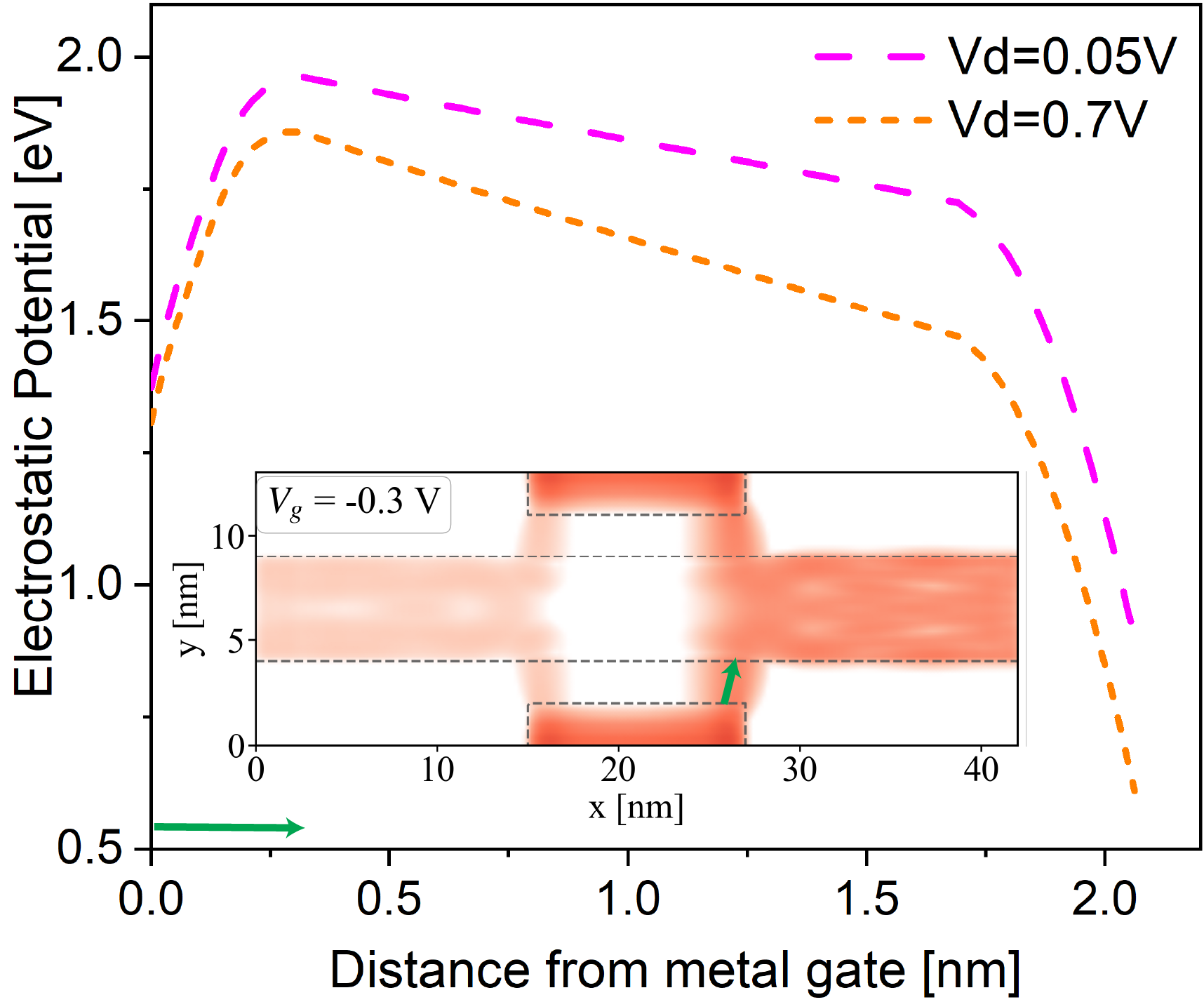}
    \caption{Effective potential profile along the dominant gate-to-drain leakage path for drain biases of $V_d=0.05$ and 0.7~V at $V_g=-0.3$~V. The potential profiles are extracted along the path indicated by the green arrow in the inset, which corresponds to the dominant leakage pathway identified from the spatial current-density map. The reduction of the high-k dielectric barrier at higher drain bias illustrates the DIDBL effect.}
    \label{fig: elec potential}
\end{figure}

The results presented in Figs.~\ref{fig: Spatial current density} and~\ref{fig: Current density streamline} identify spatially the dominant gate-drain leakage path through the dielectric at negative gate voltages. This enables us to investigate how the effective electrostatic potential along the leakage pathway evolves with drain bias and to directly visualize the DIDBL effect. Figure~\ref{fig: elec potential} shows the effective potential profile across the dielectric along the dominant leakage path identified from the current-density maps for drain biases of $V_d=0.05$ and 0.7~V. As the drain bias increases from 0.05~V to 0.7~V, the effective tunneling barrier across the dielectric is significantly reduced. This drain-induced reduction of the dielectric barrier constitutes the DIDBL effect. The reduced barrier increases the tunneling probability of carriers across the dielectric, leading to a substantial enhancement of the gate-drain leakage current for higher drain biases at negative gate voltages.

\section{Conclusion}

In this work, we demonstrated a predictive, charge-self-consistent NEGF/CBR framework\cite{Mamaluy:2026} that treats the source, drain, and gate as current-carrying quantum-mechanical contacts, enabling both a decomposition of the terminal currents and a spatially resolved map of the internal current density.

Applying this framework to a scaled GAAFET, we showed that the non-thermionic increase of the drain current at negative gate voltages originates from direct gate-to-drain tunneling through the high-$k$ dielectric, enhanced by drain-induced dielectric barrier lowering (DIDBL). The spatially resolved current density localizes this leakage at the drain-side edge of the gate and visualizes, in real space, how gate-injected carriers tunnel the dielectric and are collected by the drain, providing direct confirmation of the DIDBL-enhanced gate-drain leakage mechanism.

These results highlight the value of multi-terminal, spatially resolved quantum-transport simulations for advanced transistor design. Beyond identifying the dominant leakage path, the real-space localization of the leakage suggests possible mitigation strategies, such as locally increasing the high-$k$ dielectric thickness or employing higher-barrier dielectric materials near the drain side of the gate, without degrading gate control over the channel. Although the effectiveness of such design approaches was not evaluated in the present work and remains beyond its scope, the NEGF/CBR framework provides a predictive platform for investigating such strategies in future studies.
More broadly, the NEGF/CBR framework provides a predictive basis for identifying scaling limitations and evaluating such designs in future ultra-scaled technologies.

\begin{acknowledgments}
This work was supported in part by the U.S. Department of Energy’s Advanced Simulation and Computing (ASC) Program and the Laboratory Directed Research and Development (LDRD) program. 
J.P.M. and D.M. contributed equally to the simulation work, data analysis, interpretation of the results, and writing of the manuscript.
Sandia National Laboratories is a multimission laboratory managed and operated by National Technology and Engineering Solutions of Sandia, LLC, a wholly owned subsidiary of Honeywell International, Inc., for the U.S. Department of Energy’s National Nuclear Security Administration under contract DE-NA-0003525. This paper describes objective technical results and analysis. Any subjective views or opinions that might be expressed in the paper do not necessarily represent the views of the U.S. Department of Energy or the United States Government. This article has been authored by an employee of National Technology \& Engineering Solutions of Sandia, LLC under Contract No. DE-NA0003525 with the U.S. Department of Energy (DOE). The employee owns all right, title and interest in and to the article and is solely responsible for its contents. The United States Government retains and the publisher, by accepting the article for publication, acknowledges that the United States Government retains a non-exclusive, paid-up, irrevocable, world-wide license to publish or reproduce the published form of this article or allow others to do so, for United States Government purposes. The DOE will provide public access to these results of federally sponsored research in accordance with the DOE Public Access Plan https://www.energy.gov/downloads/doe-public-access-plan. 
\end{acknowledgments}

\section*{Data Availability}
The data that support the findings of this study are available from the corresponding author upon reasonable request.

\bibliography{reference}

\end{document}